\documentclass{article}

\usepackage{html}

\usepackage{hthtml}

\latex{\usepackage{url}}

\usepackage{graphicx}

\newcommand{\td}[1]{\mathrm{d}#1}

\newcommand{\vc}[1]{\mathbf{#1}}

\newcommand{\pnc}[1]{\textrm{#1}}
\newcommand{\br}{\nonumber\protect\\}

\html{\newcommand{\urn}[1]{\hturl{#1}}}
\latex{\newcommand{\urn}[1]{\url{#1}}}

\begin{document}

\title{Classical-Field Theory of Electron Waves as a Polarized Radiation Probe of Magnetic Surfaces}
\author{\htmladdnormallink{D.\ {}C.\
{}Hatton}{http://www.bib.hatton.btinternet.co.uk/dan/}\thanks{\htmladdnormallink{Physics
and Chemistry of Solids Group}{http://www-pcs.phy.cam.ac.uk/},
\htmladdnormallink{Cavendish Laboratory}{http://www.phy.cam.ac.uk/},
\htmladdnormallink{University of Cambridge}{http://www.cam.ac.uk/},
Madingley Road, Cambridge, UK.  CB3
0HE}\latex{\newcounter{pcs}\setcounter{pcs}{\value{footnote}}}
$^{\pnc{,}}$ \thanks{\htmladdnormallink{Girton
College}{http://www.girton.cam.ac.uk/}, \htmladdnormallink{University
of Cambridge}{http://www.cam.ac.uk/}, Huntingdon Road, Cambridge, UK.
CB3 0JG} $^{\pnc{,}}$ \thanks{\htmailto{dan.hatton@btinternet.com}} and
\htmladdnormallink{J.\ {}A.\ {}C.\
{}Bland}{http://homer.phy.cam.ac.uk/Members/Tony_Bland/Home.html}\latex{\footnotemark[\value{pcs}]}\html{\thanks{\htmladdnormallink{Physics
and Chemistry of Solids Group}{http://www-pcs.phy.cam.ac.uk/},
\htmladdnormallink{Cavendish Laboratory}{http://www.phy.cam.ac.uk/},
\htmladdnormallink{University of Cambridge}{http://www.cam.ac.uk/},
Madingley Road, Cambridge, UK.  CB3 0HE}} $^{\pnc{,}}$
\thanks{\htmladdnormallink{Selwyn College}{http://www.sel.cam.ac.uk/},
\htmladdnormallink{University of Cambridge}{http://www.cam.ac.uk/},
Grange Road, Cambridge, UK.  CB3 9DQ} $^{\pnc{,}}$
\thanks{\htmailto{jacb1@phy.cam.ac.uk}}}

\date{8th April, 2002}

\maketitle

\section*{Copyright Notice}

      Copyright \copyright\ 2002 D.\ {}C.\ {}Hatton and J.\ {}A.\ 
      {}C.\ {}Bland.  Permission is granted to copy, distribute and/or
      modify this document under the terms of the GNU Free
      Documentation License, Version 1.1 published by the Free
      Software Foundation; with no Invariant Sections, with no
      Front-Cover Texts, and with no Back-Cover Texts.  A copy of the
      license is included in the section entitled ``GNU Free
      Documentation License'' (section \ref{copying.data}.)
      
      This document can be found at
      $<$\urn{http://www.bib.hatton.btinternet.co.uk/dan/Natural_Sciences/Classical-Field_Theory_of_Electron_Waves_as_a_Polarized_Radiation_Probe_of_Magnetic_Surfaces/}$>$,
      and in transparent form at
      $<$\urn{http://www.bib.hatton.btinternet.co.uk/dan/Natural_Sciences/Classical-Field_Theory_of_Electron_Waves_as_a_Polarized_Radiation_Probe_of_Magnetic_Surfaces.tar.gz}$>$.

\section*{History}

\begin{itemize}
\item 2002, April 10th: Version 13.10, amended from version 13.4, by
  D.\ {}C.\ {}Hatton, by virtue of having arXiv file generation use
  bbl instead of bib bibliography files, and include html, hthtml,
  copyright.programs, and latex2html-copyright, and typographical
  error corrected, released.
\item 2002, April 9th: Version 13.4, amended from version 13.3, by D.\ 
  {}C.\ {}Hatton, by virtue of having typographical error corrected,
  released.
\item 2002, April 9th: Version 13.3, amended from version 12.2, by D.\ 
  {}C.\ {}Hatton, by virtue of having details of appearance of this
  document as a conference talk added, some stylistic changes which
  happened on the fly, during the talk, included, marginal notes,
  describing some things that happened during the talk, added,
  conditionality of use of url of hturl sorted out, and Makefile set
  up to produce file for arXiv, released.
\item 2002, April 4th: Version 12.2, amended, from version 11.6, by
  D.\ {}C.\ {}Hatton, by virtue of having some comments added to
  improve flow of talk, and availability of url package to \LaTeX{}
  restored, released.
\item 2002, March 29th: Version 11.6, amended, from version 11.5, by
  D.\ {}C.\ {}Hatton, by virtue of (mostly) using hthtml's hturl and
  htmailto commands instead of url's url.
\item 2002, March 28th: Version 11.5, amended, from version 11.4, by
  D.\ {}C.\ {}Hatton, by virtue of working around lack of urldef
  command in \latextohtml{}.
\item 2002, March 28th: Version 11.4, amended, from version 11.3, by
  D.\ {}C.\ {}Hatton, by virtue of having ``Conclusions'' slide added,
  ``On the Diffraction of the Magnetic Electron'' included in slide
  bibliography, clarity improved, concision improved, use of URL
  package made \LaTeX{}-only, and typographical errors corrected,
  released.
\item 2002, March 27th: Version 11.3, amended, from version 10.3, by
  D.\ {}C.\ {}Hatton, by virtue of having construction of slides
  completed, clarity improved, typographical errors corrected, and
  use of URL package introduced, released.
\item 2002, March 26th: Version 10.3, amended, from version 7.2, by
  D.\ {}C.\ {}Hatton, by virtue of having construction of slides
  begun, acknowledgements adjusted, reference to ``A Bayesian
  Perspective on Recent Epistemological Developments'' added, clarity
  improved, and typographical errors corrected, released.
\item 2002, March 25th: Version 9.1, amended, from version 7.2, by D.\ 
  {}C.\ {}Hatton, by virtue of having discussion of extension to
  multi-layers completed, philosophical commentary written, derivation
  of reflected polarization from bulk sample tidied up, and author
  file amended to work around further \latextohtml{} bugs, released.
\item 2002, March 22nd: Version 7.2, amended, from version 7.1, by D.\ 
  {}C.\ {}Hatton, by virtue of having title, author and date placed
  after $\backslash$begin\{document\}, to work around \latextohtml{}
  bug, released.
\item 2002, March 21st: Version 7.1, amended, from version 5.2, by D.\
  {}C.\ {}Hatton, by virtue of having derivation of reflected
  polarization from bulk sample completed, discussion of extension to
  multi-layers begun, Acknowledgements section added, author
  affiliations expanded and tidied up, Abstract hived off into
  separate file and formatted as section instead of as abstract
  environment, an arithmetical error corrected, and typographical
  errors corrected, released.
\item 2002, March 20th: Version 5.2, written by D.\ {}C.\ {}Hatton and
  J.\ {}A.\ {}C.\ {}Bland, released.
\end{itemize}
\section*{Endorsements}

The ``Acknowledgements'' section, and sections \ref{intro}, \ref{arc},
\ref{bulk}, \ref{multi}, \ref{incomplete}, and \ref{conclusions}, of
this document, form a transcript of a talk, given in the ``Novel
Techniques in Magnetism'' meeting of the \htmladdnormallink{CMD19CMMP
  2002}{http://physics.iop.org/IOP/Confs/CMD19/} conference, organized
by the \htmladdnormallink{European Physical
  Society}{http://www.eps.org/} and the \htmladdnormallink{Institute
  of Physics}{http://www.iop.org/}.  The ``Abstract'' is that which
appeared in the abstracts book for this conference, and on the basis
of which the talk was accepted for inclusion in the conference.

\section*{Abstract}

Recently, there has been a revival \cite{Weber:1999:EAF} of interest
\cite{Darwin:1928:DME} in mechanisms for changing the spin
polarization of an electron beam on transmission through, or
reflection from, a magnetic surface.  An understanding of these
mechanisms would \cite{Weber:1999:EAF} allow the use of an electron
beam as a polarized radiation probe for magnetic characterization,
like light in MOKE and neutrons in PNR.  Here, a mechanism is
described which, unlike simultaneously occurring processes proposed
elsewhere \cite{Weber:1999:EAF}, polarizes an unpolarized incident
beam without recourse to inelastic processes.

\begin{figure}
\includegraphics[width=\textwidth, height=0.75\textwidth]{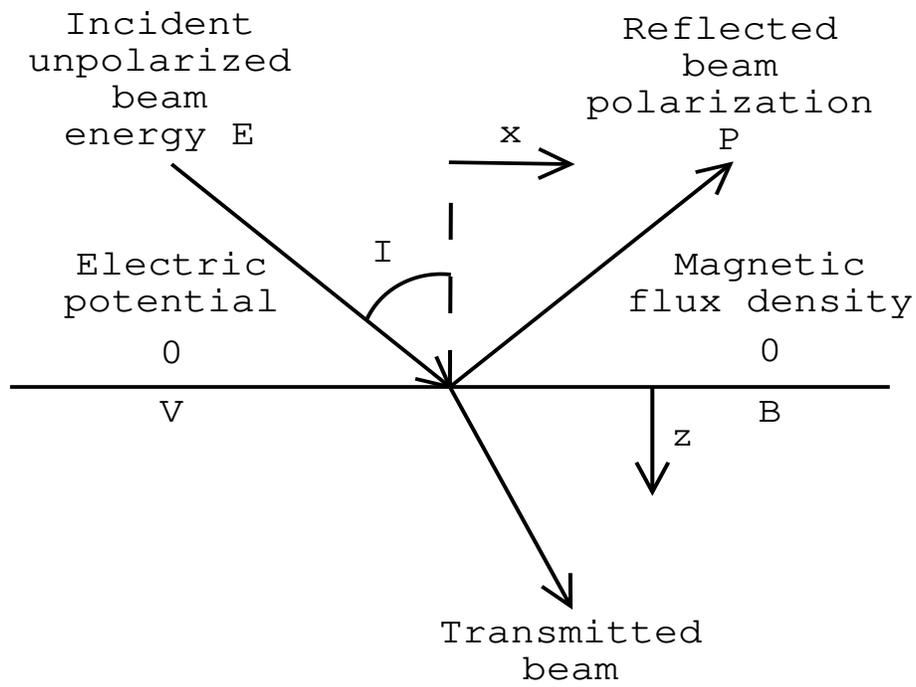}
\caption{Surface of a Bulk Magnetic Sample}
\label{dhatton1}
\end{figure}

A magnetic field leads to a Zeeman term in an electron's Hamiltonian,
which depends on the angle $\theta$ between the electron's spin vector
and the magnetic flux.  As a result, when an electron wave is incident
on the surface of a bulk magnetic material (figure \ref{dhatton1},)
the wave-number of the transmitted wave depends on $\theta$.  When the
conditions of continuity of the wave-function, and of its first
spatial derivative, at the surface, and conservation of particles, are
applied, an electron reflection coefficient is obtained which also
depends on $\theta$.  Therefore, some polarizations are preferentially
reflected, while others are preferentially transmitted.  The amplitude
reflection and transmission coefficients can readily be converted to
intensity coefficients, and averaged over an incoherent superposition
of electron waves of different $\theta$, e.g.\ {}an unpolarized
incident beam.  The reflected polarization is
\begin{equation}
P = -\frac{2e\mu_BVB}{3e^2V^2+\mu_B^2B^2}\pnc{,}
\end{equation}
which can take values
\begin{equation}
-\frac{1}{\sqrt{3}}\leq{}P\leq\frac{1}{\sqrt{3}}\pnc{,}
\end{equation}
depending on the balance between $V$ and $B$.

The analysis can be extended to multi-layers using the theory of
Fabry-Perot etalons.

\section*{Acknowledgements}

Good morning.  As you've just heard, I'm Dan Hatton, and I'm going to
present a classical-field theory of electron waves as a polarized
radiation probe of magnetic surfaces.

We'd like to thank the Engineering and Physical Sciences Research
Council, in the UK, for paying for this work, and also Thanos
Mitrelias, Klaus Peter Kopper, and Peter Bode, our present and past
collaborators, in setting up the experiments for which we're putting
forward a theoretical interpretation.

\section{Introduction}
\label{intro}
\subsection{Polarized Neutron Reflection}

Polarized neutron reflection, or PNR, is \cite{Blundell:1992:PNR,
  Blundell:1993:PNR} an established experimental technique for the
measurement of layer-dependent magnetization vector in magnetic
multi-layers.  A multi-layer structure is \cite{Blundell:1992:PNR,
  Blundell:1993:PNR} modelled as a series of steps in nuclear
potential and magnetic flux density.  The amplitude reflection
coefficient for neutron waves at each step is then calculated by
applying the usual \cite{Rae:1992:QM} boundary conditions to the
spin-up and spin-down wave-functions at the step, given the change in
wave-vector produced by the potential step.  The change in wave-vector
depends on the neutron's spin direction because of the torque exerted
upon the neutron magnetic moment, by the magnetic field.  Therefore,
the spin polarization of the reflected neutron beam, as a function of
incident beam energy, provides an indicator of the depth-resolved
magnetization profile of the sample.

\subsection{Polarized Electron Reflection}

Polarized electron reflection and diffraction are \cite{Lind:1986:ETC,
  Lind:1987:SSA} also established experimental techniques, for the
characterization of magnetic surfaces.  The measurement is identical
to PNR except for the substitution of electrons for neutrons, and the
unavailability \cite{Kessler:1976:PE, Kessler:1985:PE} of the
Stern-Gerlach experiment, either for controlling the incident
polarization, or for measuring the reflected polarization.  The
Stern-Gerlach experiment is \cite{Lind:1986:ETC, Lind:1987:SSA}
typically replaced by a Mott polarimeter \cite{Hodge:1979:MES,
  Gay:1992:MEP, Dunning:1994:MEP}, for measuring the reflected
polarization.  Electrons have significant advantages over neutrons for
this purpose: an electron beam can be produced using a device roughly
equivalent to a light-bulb filament, whereas a neutron beam is
typically produced using a nuclear reactor.  Also, the magnetic moment
of the electron is nearly two thousand times that of the neutron.

Despite the long-standing use of polarized electron reflection as an
experimental technique, as far as we're aware, there has been no
attempt to develop a theoretical model of the process, along the lines
of that used for PNR, in order to interpret the results in terms of
the depth profile of the magnetization in the sample.  Our intention
here is to produce an analysis of polarized electron reflection
similar to that of PNR by Blundell and Bland \cite{Blundell:1992:PNR,
  Blundell:1993:PNR}.  If you find the analysis interesting, a
transcript of this talk can be found on the web at
\htmladdnormallink{this
  address}{http://www.bib.hatton.btinternet.co.uk/dan/Natural_Sciences/Classical-Field_Theory_of_Electron_Waves_as_a_Polarized_Radiation_Probe_of_Magnetic_Surfaces/}.
The web version also includes more details of the derivations of
equations, which are only sketched here.  If you can stomach reading
all that maths, we'd be grateful for any comments or suggestions.  If
you can't stomach reading all that maths, I suggest you make yourself
difficult to contact around October, so you don't end up being one of
the unfortunates who have to examine my thesis.

\section{Amplitude Reflection Coefficient for an Electron Pure State,
at a Single Step in Electric Potential and Magnetic Flux Density}
\label{arc}

\begin{figure}
\includegraphics[width=\textwidth, height=0.75\textwidth]{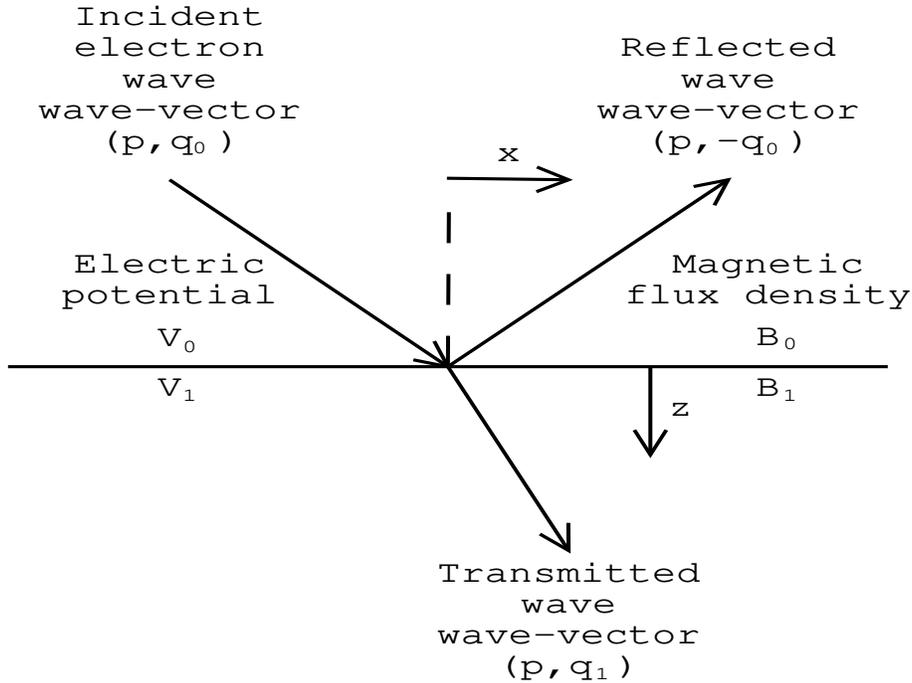}
\caption{Reflection of an Electron Wave by a Single Step in Electric
Potential and Magnetic Flux Density}
\label{single_step}
\end{figure}
The first step in the analysis of reflection is to build a
potential-theory model of the sample, as a series of steps in electric
potential and magnetic flux density (figure \ref{dhatton1}, figure
\ref{single_step}.)  Next, we need to discover the amplitude
reflection coefficient, for a pure, coherent, electron wave, at a
single step (figure \ref{single_step}.)  The incident and transmitted
electron waves are modelled as plane waves, with well-defined
wave-vector components $p$ in the plane of the interface, and $q_i$
perpendicular to the interface.  $p$ must be the same for all the
waves, in order to satisfy the boundary condition of continuity of the
wave-function at the interface.  Strictly, the eigen-states of a
Hamiltonian which includes a magnetic field are not plane waves; more
about this later (section \ref{incomplete}.)  The amplitude reflection
coefficient is \cite{Rae:1992:QM} this:
\begin{equation}
r_{01} = \frac{q_0-q_1}{q_0+q_1}\pnc{,}
\end{equation}
or, for a general interface, this one:
\begin{equation}
r_{ij} = \frac{q_i-q_j}{q_i+q_j}\pnc{.}
\end{equation}

Next, we need to build an expression for the energy of the electrons.
There will be kinetic energy terms, along with an electrostatic
potential energy, and a term due to the torque, on the electron
magnetic moment, in a magnetic field \cite{Rae:1992:QM}. The form used
for this last term assumes a well-defined energy for all directions of
the electron spin.  Strictly, only certain spin directions are
eigen-states of a Hamiltonian which includes a magnetic field; more
about this later (section \ref{incomplete}.)  This leads to this
expression
\begin{equation}
q_i =
\left(\frac{2m_eE\cos^2I}{\hbar^2}\right)^{1/2}(1+x_i)^{1/2}\pnc{,}
\end{equation}
for the perpendicular wave-vector component, where $I$ represents an
angle of incidence, and the potential energy terms are represented by
these dimensionless numbers:
\begin{equation}
x_i = y_i+z_i\cos{}S_i\pnc{,}
\end{equation}
\begin{equation}
y_i = \frac{eV_i}{E\cos^2I}\pnc{,}
\end{equation}
\begin{equation}
z_i = -\frac{e\hbar{}B_i}{2m_eE\cos^2I}\pnc{;}
\end{equation}
I have a big enough ego to call them the Hatton numbers, but I suspect
I wouldn't get away with it.  $S_i$ is the angle between the electron
spin direction and the magnetic flux density in region $i$, and $E$ is
the total energy of the incident electrons, and therefore, by
conservation of energy, of all the electrons.

We now use a binomial expansion \cite{Gallagher::MFH} for the case
where the potential energy terms are much smaller than the total
electron energy, where the dimensionless numbers we've just devised
are small.  The magnetic term associated with the Weiss field in a
ferromagnet is \cite{Moruzzi:1978:CEP} a few tenths of an
electron-volt, and the electrostatic contact potentials in the metals
which we study will not be more than a few volts\marginpar{In the
  conference talk, I made an error, and had to correct myself, here.
  Only the second, corrected version appears in this document.},
whereas, in our experimental set-up, the incident electron energies
range from a few hundred to a few thousand electron volts, so this
approximation seems reasonable.  With this expansion, the amplitude
reflection coefficient is this:
\begin{equation}
r_{ij} = \frac{1}{4}x_i-\frac{1}{4}x_j-\frac{1}{8}x_i^2-\frac{1}{8}x_ix_j+\frac{1}{8}x_j^2+O(\{x_i,x_j\}^3)\pnc{.}
\end{equation}

\section{Reflection of an Unpolarized Beam from the Surface of a Bulk
  Magnetic Sample}
\label{bulk}

An unpolarized incident electron beam is \cite{Kessler:1976:PE,
  Kessler:1985:PE} an incoherent superposition of pure states
representing all directions of the incident spin.  The polarization of
the reflected beam from any surface is, therefore, given by an average
of the polarization over all polarization directions, weighted
according to the intensity reflection coefficient for each
polarization.  This incoherent averaging process (section
\ref{bulk-details}) gives this reflected polarization from a bulk surface
(figure \ref{dhatton1})
\begin{eqnarray}
P & = & \frac{2y_1z_1}{3y_1^2+z_1^2}+O(\{y_1,z_1\})\br
& = &
-\frac{4e^2\hbar{}m_eV_1B_1}{12e^2m_e^2V_1^2+e^2\hbar^2B_1^2}+O(\{y_1,z_1\})\pnc{.}
\end{eqnarray}
Both the term given explicitly, and the next term in the binomial
expansion, are in the direction of the magnetic flux density in the
bulk material.

The most salient qualitative feature of this polarization formula is
that, at high incident electron energies, the reflected polarization
is dominated by a non-zero term, which is independent of the incident
electron energy, and controlled by the balance between the
electrostatic potential and the magnetic flux density, in the sample.
This polarization can be as large as $\frac{1}{\sqrt{3}}$ in either
direction.

\section{Multi-Layer Structures}
\label{multi}

\begin{figure}
\includegraphics[width=\textwidth, height=0.75\textwidth]{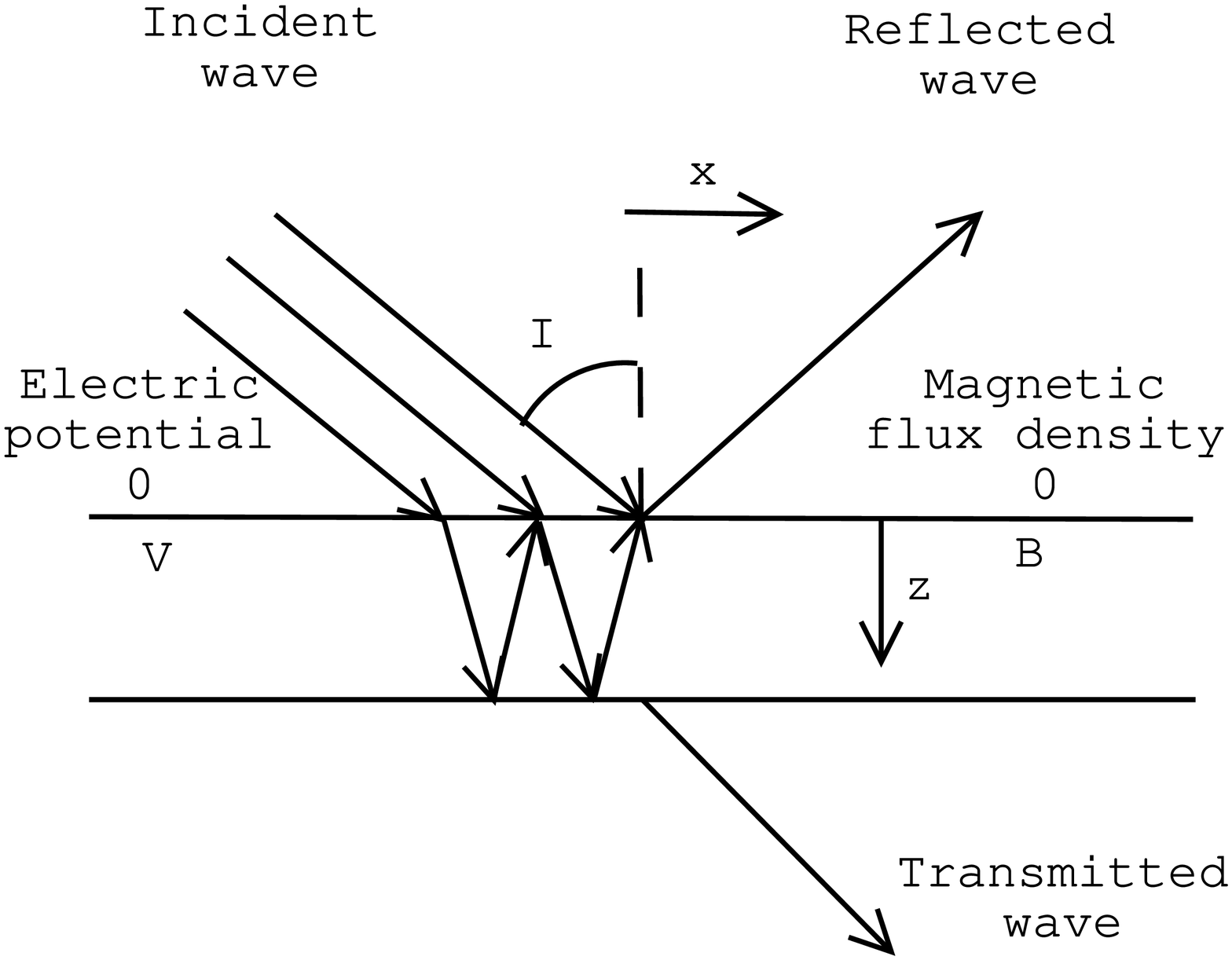}
\caption{Reflection Paths for an Electron Wave in a Single Magnetic Layer}
\label{single-layer}
\end{figure}
We propose to extend this analysis to magnetic multi-layer structures,
by using the theory of Fabry-Perot etalons, as is
\cite{Blundell:1992:PNR, Blundell:1993:PNR} already the practice in
PNR.  There are infinitely many possible paths for reflection from a
multi-layer structure, indexed by how many times the electron wave
``bounces'' within each layer.  In the diagram (figure
\ref{single-layer},) we can see paths with no bounces, with one
bounce, and with two bounces.  For a given, pure incident wave, the
reflected waves from the various paths are superposed coherently to
build the reflected wave, each term in the coherent superposition
including an amplitude factor due to the amplitude reflection or
transmission coefficient at each interface which it has
encountered\marginpar{In the conference talk, I made an error, and had
  to correct myself, here.  Only the second, corrected version appears
  in this document.}, and a phase factor due to the path length which
it has traversed in the magnetic layers.  This will result in a
spin-dependent amplitude reflection coefficient for the whole
multi-layer system, which will provide the weightings to go into the
incoherent superposition over an unpolarized incident beam.  This
incoherent superposition, as for the bulk sample, will give the
reflected polarization.  We expect working through the maths for this
to be trivial, but time-consuming.

\section{Comments on This Analysis}
\label{incomplete}

I promised to comment on some assumptions in this analysis.  Firstly,
there's the matter of modelling the electrons as a plane wave.  This
is equivalent to neglecting the deflection of the electrons by the
Lorentz force, which means taking the limit of weak magnetic fields;
something we've done in the binomial expansion anyway.  The same
convention of neglecting this deflection was adopted by Weber \emph{et
  al.} \cite{Weber:1999:EAF}, when they analysed the spin polarization
of transmitted electron waves.

Secondly, there's the issue of pretending that all electron spin
directions are eigen-states of the Hamiltonian.  In this we depart
from the tradition of analysis of PNR, where matrices are
\cite{Blundell:1992:PNR, Blundell:1993:PNR} used to represent the
Zeeman energy, and the reflection coefficient, without any need for
this approximation.  We also depart from the work of Weber \emph{et
  al.} \cite{Weber:1999:EAF} on electron transmission: they regard the
Larmor precession, which is a manifestation of the fact that not all
spin directions are eigen-states of the Hamiltonian in a magnetic
field, as crucial in determining the transmitted polarization.  We
intend to produce a more ``first-principles'' model in the near
future, which will use the matrix representation of the reflection
coefficients, and therefore capture the Larmor precession, and other
spin-flip scattering effects.  However, we don't intend to devise this
model as a replacement for the one presented here, but as a complement
to it.  What we'd like to do is subject both models, along with a
third, completely classical, reflection model, to experimental data,
and use the well-established \cite{MacKay:1992:BMA} methods of
Bayesian statistics, first to infer the parameters of magnetic flux
density, electric potential, and layer thickness, for each model, then
to judge the relative confidence which we have in each model.

One reason for not simply abandoning all but the most
``first-principles'' of the models is given by Anderson
\cite{Anderson:1972:MID}, who points out that any system, more
complicated than a molecule of four atoms or so, is pretty well never
in an eigen-state of its Hamiltonian, so the Schr\"odinger equation
doesn't describe the state of the system.  This is because the
tunnelling-like processes, which would otherwise collapse the system
into an eigen-state of its Hamiltonian, are very slow for complicated
systems: often very slow compared with the age of the universe, and
certainly very slow compared with the rate of occurrence of
measurement-like interactions with the outside world, which collapse
the system into eigen-states of operators other than the Hamiltonian.
Therefore, it can't be guaranteed that the model which implements a
Schr\"odinger equation with the most realistic Hamiltonian will always
be the most useful in describing the real behaviour of the system.

I might be inclined to add to this a very different argument
\cite{Hatton:2002:BPR} for not always preferring the most
first-principles model, but this isn't the time or the place for my
speculations on mathematical philosophy.  Anyone who has a burning
desire to hear them can find them via the reference on the slide.

Thirdly, it's worth commenting on the effect on the polarization of
transmitted waves, due to spin-dependent loss of electrons to
inelastic processes, which was noted by Weber \emph{et al.}
\cite{Weber:1999:EAF}.  At first glance, our classical-field analysis
appears to be entirely elastic.  However, it is capable of
assimilating the effect of these processes, which will appear as
imaginary parts in the electric potential and magnetic flux density.

\section{Conclusions}
\label{conclusions}

OK.  What have we learnt?

\begin{itemize}
\item The spin polarization of the reflected electron beam from a bulk
  magnetic surface, in the model described, is this:
\begin{equation}
\vc{P} = -\frac{4e^2\hbar{}m_eV_1\vc{B_1}}{12e^2m_e^2V_1^2+e^2\hbar^2B_1^2}+O(\{y_1,z_1\})\pnc{.}
\end{equation}
\item The extension of the model to multi-layer systems is likely to
  be a trivial, but time-consuming, mathematical task.
\item Two other, similar models are planned, one of which differs from
  this by the use of a more ``first-principles'' treatment of the
  Zeeman energy, and the other by a fully classical treatment of the
  reflection process, and
\item there is a strong case for retaining all three models, and using
  Bayesian statistics to compare them in the light of experimental
  data, rather than discarding all but the most ``first-principles''
  model.
\end{itemize}

Thank you for listening.  I'll show the slides of references
gradually, while I field some questions\marginpar{This actually failed
  to happen, except for the first four references: answering the
  questions kept me too busy to change over the slides.  If you're
  reading this document because you wanted to know about the
  references, you're in luck.  They're here, and numbered in the same
  way as in the slides.}.

\section{Supplementary Mathematical Details}
\subsection{Amplitude Reflection Coefficient for an Electron Pure State,
  at a Single Step in Electric Potential and Magnetic Flux Density}

The incident and transmitted electron waves are modelled (figure
\ref{single_step}) as plane waves, allowing the well-defined
wave-vector components $p$ in the plane of the interface, and $q_i$
perpendicular to the interface.  $p$ must be the same for all the
waves, in order to satisfy the boundary condition of continuity of the
wave-function at the interface.  The amplitude reflection coefficient
is \cite{Rae:1992:QM}
\begin{equation}
r_{01} = \frac{q_0-q_1}{q_0+q_1}\pnc{,}
\end{equation}
or, for a general interface,
\begin{equation}
r_{ij} = \frac{q_i-q_j}{q_i+q_j}\pnc{.}
\end{equation}

Next, we need to build an expression for the energy of the electrons.
There will be kinetic energy terms, which, in the non-relativistic
limit, are
\[
\frac{\hbar^2p^2}{2m_e}\pnc{,}
\]
and
\[
\frac{\hbar^2q_i^2}{2m_e}\pnc{,}
\]
along with an electrostatic potential energy
\[
-eV_i\pnc{,}
\]
and a term due to the torque, on the electron magnetic moment, in a
magnetic field \cite{Rae:1992:QM}
\[
\frac{e\hbar{}B_i\cos{}S_i}{2m_e}\pnc{,}
\]
where $S_i$ is the angle between the electron spin and the magnetic
flux density.  The form of this last term assumes a well-defined
energy for all values of $S_i$.  Strictly, only certain $S_i$ values
are eigen-states of a Hamiltonian which includes a magnetic field;
more about this later (section \ref{incomplete}.)  The total energy is
\begin{equation}
E =
\frac{\hbar^2p^2}{2m_e}+\frac{\hbar^2q_i^2}{2m_e}-eV_i+\frac{e\hbar{}B_i\cos{}S_i}{2m_e}\pnc{,}
\end{equation}
or, where $p$ is expressed as a fraction $\sin{}I$ of the total wave-number
in the absence of any potential, $I$ being an angle of incidence like
that in figure \ref{dhatton1},
\begin{equation}
E = E\sin^2I+\frac{\hbar^2q_i^2}{2m_e}-eV_i+\frac{e\hbar{}B_i\cos{}S_i}{2m_e}
\end{equation}
\begin{eqnarray}
\Rightarrow{}q_i & = &
\left(\frac{2m_eE\cos^2I}{\hbar^2}\right)^{1/2}\left(1+\frac{eV_i}{E\cos^2I}-\frac{e\hbar{}B_i\cos{}S_i}{2m_eE\cos^2I}\right)^{1/2}\br
& = & \left(\frac{2m_eE\cos^2I}{\hbar^2}\right)^{1/2}(1+x_i)^{1/2}\pnc{,}
\end{eqnarray}
where $x_i = y_i+z_i\cos{}S_i$, $y_i = \frac{eV_i}{E\cos^2I}$, and
$z_i = -\frac{e\hbar{}B_i}{2m_eE\cos^2I}$.

We now use a binomial expansion \cite{Gallagher::MFH} for the case
where the potential energy terms are much smaller than the total
electron energy, where the dimensionless numbers we've just devised
are small.  The magnetic term associated with the Weiss field in a
ferromagnet is \cite{Moruzzi:1978:CEP} a few tenths of an
electron-volt, and the electrostatic contact potentials in the metals
which we study will not be more than a few volts, whereas, in our
experimental set-up, the incident electron energies range from a few
hundred to a few thousand electron volts, so this approximation seems
reasonable.
\begin{eqnarray}
q_i & = & \left(\frac{2m_eE\cos^2I}{\hbar^2}\right)^{1/2}\left(1+\frac{1}{2}x_i-\frac{1}{8}x_i^2+O(x_i^3)\right)\pnc{.}
\end{eqnarray}

The amplitude reflection coefficient is, therefore,
\begin{eqnarray}
r_{ij} & = &
\frac{1}{2}\left(\frac{1}{2}x_i-\frac{1}{2}x_j+\frac{1}{8}x_j^2-\frac{1}{8}x_i^2+O(\{x_i,x_j\}^3)\right)\left(1+\frac{1}{4}x_i+\frac{1}{4}x_j-\frac{1}{16}x_i^2-\frac{1}{16}x_j^2+O(\{x_i,x_j\}^3)\right)^{-1}\br
& = &
\frac{1}{2}\left(\frac{1}{2}x_i-\frac{1}{2}x_j+\frac{1}{8}x_j^2-\frac{1}{8}x_i^2+O(\{x_i,x_j\}^3)\right)\left(1-\frac{1}{4}x_i-\frac{1}{4}x_j+\frac{1}{8}x_i^2+\frac{1}{8}x_ix_j+\frac{1}{8}x_j^2+O(\{x_i,x_j\}^3)\right)\br
& = &
\frac{1}{2}\left(\frac{1}{2}x_i-\frac{1}{2}x_j-\frac{1}{4}x_i^2-\frac{1}{4}x_ix_j+\frac{1}{4}x_j^2+O(\{x_i,x_j\}^3)\right)\br
& = & \frac{1}{4}x_i-\frac{1}{4}x_j-\frac{1}{8}x_i^2-\frac{1}{8}x_ix_j+\frac{1}{8}x_j^2+O(\{x_i,x_j\}^3)\pnc{.}
\end{eqnarray}

\subsection{Reflection of an Unpolarized Beam from the Surface of a Bulk
  Magnetic Sample}
\label{bulk-details}

An unpolarized incident electron beam is \cite{Kessler:1976:PE,
  Kessler:1985:PE} an incoherent superposition of pure states
  representing all directions of the incident spin.  Each such
  direction can be represented by its spherical polar angle
  co-ordinates $(\theta{},\phi{})$.  That is to say, the incident beam
  contains a flux of electrons
\begin{equation}
F_1\td{\theta}\td{\phi} = A\sin\theta\td{\theta}\td{\phi}
\end{equation}
with polarization direction between $\theta$ and
$\theta{}+\td{\theta}$, and between $\phi$ and $\phi{}+\td{\phi}$.
The flux of such electrons in the reflected beam will, therefore, be
\begin{equation}
F_2\td{\theta}\td{\phi} = |r_{ij}|^2F_1\td{\theta}\td{\phi}\pnc{.}
\end{equation}

The reflection from the surface of a bulk sample is to be modelled as
a single reflection, of amplitude reflection coefficient $r_{01}$, in
a situation where $V_0$, $B_0$, and therefore $x_0$, are all zero.  In
this case,
\begin{equation}
r_{01} = -\frac{1}{4}x_1+\frac{1}{8}x_1^2+O(x_1^3)\pnc{,}
\end{equation}
and
\begin{equation}
|r_{01}|^2 = \frac{1}{16}x_1^2-\frac{1}{16}x_1^3+O(x_1^4)\pnc{,}
\end{equation}
assuming that $r_{01}$ is real.

If the spherical polar representation $(\theta_i,\phi_i)$ is used for
the direction of the magnetic flux density in region $i$, then
\begin{equation}
\cos{}S_i = \sin\theta_i\cos\phi_i\sin\theta\cos\phi{}+\sin\theta_i\sin\phi_i\sin\theta\sin\phi{}+\cos\theta_i\cos\theta\pnc{.}
\end{equation}
Therefore,
\begin{equation}
x_i = y_i+z_i(\sin\theta_i\cos\phi_i\sin\theta\cos\phi{}+\sin\theta_i\sin\phi_i\sin\theta\sin\phi{}+\cos\theta_i\cos\theta{})\pnc{.}
\end{equation}

The polarization of the pure state represented by $(\theta{},\phi{})$,
in the Cartesian co-ordinate system associated with this spherical
polar system, is
\begin{equation}
\vc{P}(\theta{},\phi{}) =
(\sin\theta\cos\phi{},\sin\theta\sin\phi{},\cos\theta{})\pnc{,}
\end{equation}
and the average polarization of the reflected beam is
\begin{eqnarray}
\vc{P} & = &
\frac{\int_{\theta{}=0}^{\pi}\int_{\phi=0}^{2\pi}\vc{P}(\theta{},\phi{})F_2\td{\theta}\td{\phi}}{\int_{\theta{}=0}^{\pi}\int_{\phi=0}^{2\pi}F_2\td{\theta}\td{\phi}}\br
& = &
\frac{\int_{\theta{}=0}^{\pi}\int_{\phi=0}^{2\pi}(\sin\theta\cos\phi{},\sin\theta\sin\phi{},\cos\theta{})|r_{01}|^2F_1\td{\theta}\td{\phi}}{\int_{\theta{}=0}^{\pi}\int_{\phi=0}^{2\pi}|r_{01}|^2F_1\td{\theta}\td{\phi}}\br
& = &
\frac{\int_{\theta{}=0}^{\pi}\int_{\phi=0}^{2\pi}(\sin\theta\cos\phi{},\sin\theta\sin\phi{},\cos\theta{})\left(\frac{1}{16}x_1^2-\frac{1}{16}x_1^3+O(x_1^4)\right)\sin\theta\td{\theta}\td{\phi}}{\int_{\theta{}=0}^{\pi}\int_{\phi=0}^{2\pi}\left(\frac{1}{16}x_1^2-\frac{1}{16}x_1^3+O(x_1^4)\right)\sin\theta\td{\theta}\td{\phi}}\br
& = & \frac{(I_2-I_6,I_3-I_7,I_4-I_8)+O(\{y_1,z_1\}^4)}{I_1-I_5+O(\{y_1,z_1\}^4)}\pnc{.}
\end{eqnarray}

The crucial integrals are
\begin{eqnarray}
I_1 & = & \int_{\theta{}=0}^{\pi}\int_{\phi=0}^{2\pi}x_1^2\sin\theta\td{\theta}\td{\phi}\br
& = &
\int_{\theta{}=0}^{\pi}\int_{\phi=0}^{2\pi}(y_1+z_1(\sin\theta_1\cos\phi_1\sin\theta\cos\phi{}+\sin\theta_1\sin\phi_1\sin\theta\sin\phi{}+\cos\theta_1\cos\theta{}))^2\sin\theta\td{\theta}\td{\phi}\br
& = &
\int_{\theta{}=0}^{\pi}\int_{\phi=0}^{2\pi}(y_1^2\sin\theta{}\br
& & +2y_1z_1(\sin\theta_1\cos\phi_1\sin^2\theta\cos\phi{}+\sin\theta_1\sin\phi_1\sin^2\theta\sin\phi{}+\cos\theta_1\sin\theta\cos\theta{})\br
& &
+z_1^2(\sin^2\theta_1\cos^2\phi_1\sin^3\theta\cos^2\phi{}+2\sin^2\theta_1\sin\phi_1\cos\phi_1\sin^3\theta\sin\phi{}\cos\phi{}\br
& &
+2\sin\theta_1\cos\theta_1\cos\phi_1\sin^2\theta\cos\theta\cos\phi{}+\sin^2\theta_1\sin^2\phi_1\sin^3\theta\sin^2\phi\br
& & +2\sin\theta_1\cos\theta_1\sin\phi_1\sin^2\theta\cos\theta\sin\phi+\cos^2\theta_1\sin\theta\cos^2\theta{}))\td{\theta}\td{\phi}\br
& = & 4\pi{}y_1^2+\frac{4\pi{}z_1^2}{3}\pnc{,}
\end{eqnarray}
\begin{eqnarray}
I_2 & = & \int_{\theta{}=0}^{\pi}\int_{\phi=0}^{2\pi}x_1^2\sin^2\theta\cos\phi\td{\theta}\td{\phi}\br
& = & \int_{\theta{}=0}^{\pi}\int_{\phi=0}^{2\pi}(y_1+z_1(\sin\theta_1\cos\phi_1\sin\theta\cos\phi{}+\sin\theta_1\sin\phi_1\sin\theta\sin\phi{}+\cos\theta_1\cos\theta{}))^2\sin^2\theta\cos\phi\td{\theta}\td{\phi}\br
& = &
\int_{\theta{}=0}^{\pi}\int_{\phi=0}^{2\pi}(y_1^2\sin^2\theta{}\cos\phi{}\br
& & +2y_1z_1(\sin\theta_1\cos\phi_1\sin^3\theta\cos^2\phi{}+\sin\theta_1\sin\phi_1\sin^3\theta\sin\phi{}\cos\phi{}+\cos\theta_1\sin^2\theta\cos\theta{}\cos\phi{})\br
& &
+z_1^2(\sin^2\theta_1\cos^2\phi_1\sin^4\theta\cos^3\phi{}+2\sin^2\theta_1\sin\phi_1\cos\phi_1\sin^4\theta\sin\phi{}\cos^2\phi{}\br
& &
+2\sin\theta_1\cos\theta_1\cos\phi_1\sin^3\theta\cos\theta\cos^2\phi{}+\sin^2\theta_1\sin^2\phi_1\sin^4\theta\sin^2\phi\cos\phi\br
& &
+2\sin\theta_1\cos\theta_1\sin\phi_1\sin^3\theta\cos\theta\sin\phi\cos\phi{}+\cos^2\theta_1\sin^2\theta\cos^2\theta\cos\phi))\td{\theta}\td{\phi}\br
& = & \frac{8\pi{}y_1z_1\sin\theta_1\cos\phi_1}{3}
\pnc{,}
\end{eqnarray}
\begin{eqnarray}
I_3 & = & \int_{\theta{}=0}^{\pi}\int_{\phi=0}^{2\pi}x_1^2\sin^2\theta\sin\phi\td{\theta}\td{\phi}\br
& = &
\int_{\theta{}=0}^{\pi}\int_{\phi=0}^{2\pi}(y_1+z_1(\sin\theta_1\cos\phi_1\sin\theta\cos\phi{}+\sin\theta_1\sin\phi_1\sin\theta\sin\phi{}+\cos\theta_1\cos\theta{}))^2\sin^2\theta\sin\phi\td{\theta}\td{\phi}\br
& = &
\int_{\theta{}=0}^{\pi}\int_{\phi=0}^{2\pi}(y_1^2\sin^2\theta{}\sin\phi\br
& & +2y_1z_1(\sin\theta_1\cos\phi_1\sin^3\theta\sin\phi\cos\phi{}+\sin\theta_1\sin\phi_1\sin^3\theta\sin^2\phi{}+\cos\theta_1\sin^2\theta\cos\theta\sin\phi{})\br
& &
+z_1^2(\sin^2\theta_1\cos^2\phi_1\sin^4\theta\sin\phi\cos^2\phi{}+2\sin^2\theta_1\sin\phi_1\cos\phi_1\sin^4\theta\sin^2\phi{}\cos\phi{}\br
& &
+2\sin\theta_1\cos\theta_1\cos\phi_1\sin^3\theta\cos\theta\sin\phi\cos\phi{}+\sin^2\theta_1\sin^2\phi_1\sin^4\theta\sin^3\phi\br
& &
+2\sin\theta_1\cos\theta_1\sin\phi_1\sin^3\theta\cos\theta\sin^2\phi+\cos^2\theta_1\sin^2\theta\cos^2\theta\sin\phi{}))\td{\theta}\td{\phi}\br
& = & \frac{8\pi{}y_1z_1\sin\theta_1\sin\phi_1}{3}
\pnc{,}
\end{eqnarray}
\begin{eqnarray}
I_4 & = &
\int_{\theta{}=0}^{\pi}\int_{\phi=0}^{2\pi}x_1^2\sin\theta\cos\theta\td{\theta}\td{\phi}\br
& = &
\int_{\theta{}=0}^{\pi}\int_{\phi=0}^{2\pi}(y_1+z_1(\sin\theta_1\cos\phi_1\sin\theta\cos\phi{}+\sin\theta_1\sin\phi_1\sin\theta\sin\phi{}+\cos\theta_1\cos\theta{}))^2\sin\theta\cos\theta\td{\theta}\td{\phi}\br
& = &
\int_{\theta{}=0}^{\pi}\int_{\phi=0}^{2\pi}(y_1^2\sin\theta\cos\theta{}\br
& & +2y_1z_1(\sin\theta_1\cos\phi_1\sin^2\theta\cos\theta\cos\phi{}+\sin\theta_1\sin\phi_1\sin^2\theta\cos\theta\sin\phi{}+\cos\theta_1\sin\theta\cos^2\theta{})\br
& &
+z_1^2(\sin^2\theta_1\cos^2\phi_1\sin^3\theta\cos\theta\cos^2\phi{}+2\sin^2\theta_1\sin\phi_1\cos\phi_1\sin^3\theta\cos\theta\sin\phi{}\cos\phi{}\br
& &
+2\sin\theta_1\cos\theta_1\cos\phi_1\sin^2\theta\cos^2\theta\cos\phi{}+\sin^2\theta_1\sin^2\phi_1\sin^3\theta\cos\theta\sin^2\phi\br
& &
+2\sin\theta_1\cos\theta_1\sin\phi_1\sin^2\theta\cos^2\theta\sin\phi+\cos^2\theta_1\sin\theta\cos^3\theta{}))\td{\theta}\td{\phi}\br
& = & \frac{8\pi{}y_1z_1\cos\theta_1}{3}\pnc{,}
\end{eqnarray}
\begin{eqnarray}
I_5 & = &
\int_{\theta{}=0}^{\pi}\int_{\phi=0}^{2\pi}x_1^3\sin\theta\td{\theta}\td{\phi}\br
& = &
\int_{\theta{}=0}^{\pi}\int_{\phi=0}^{2\pi}(y_1+z_1(\sin\theta_1\cos\phi_1\sin\theta\cos\phi{}+\sin\theta_1\sin\phi_1\sin\theta\sin\phi{}+\cos\theta_1\cos\theta{}))^3\sin\theta\td{\theta}\td{\phi}\br
& = &
\int_{\theta{}=0}^{\pi}\int_{\phi=0}^{2\pi}(y_1^3\sin\theta\br
& & +3y_1^2z_1(\sin\theta_1\cos\phi_1\sin^2\theta\cos\phi{}+\sin\theta_1\sin\phi_1\sin^2\theta\sin\phi{}+\cos\theta_1\sin\theta\cos\theta{})\br
& & +3y_1z_1^2(\sin^2\theta_1\cos^2\phi_1\sin^3\theta\cos^2\phi{}+2\sin^2\theta_1\sin\phi_1\cos\phi_1\sin^3\theta\sin\phi{}\cos\phi{}\br
& &
+2\sin\theta_1\cos\theta_1\cos\phi_1\sin^2\theta\cos\theta\cos\phi{}+\sin^2\theta_1\sin^2\phi_1\sin^3\theta\sin^2\phi\br
& &
+2\sin\theta_1\cos\theta_1\sin\phi_1\sin^2\theta\cos\theta\sin\phi{}+\cos^2\theta_1\sin\theta\cos^2\theta{})\br
& &
+z_1^3(\sin^3\theta_1\cos^3\phi_1\sin^4\theta\cos^3\phi{}+3\sin^3\theta_1\sin\phi_1\cos^2\phi_1\sin^4\theta\sin\phi\cos^2\phi{}\br
& &
+3\sin^2\theta_1\cos\theta_1\cos^2\phi_1\sin^3\theta\cos\theta\cos^2\phi{}+3\sin^3\theta_1\sin^2\phi_1\cos\phi_1\sin^4\theta\sin^2\phi\cos\phi{}\br
& &
+6\sin^2\theta_1\cos\theta_1\sin\phi_1\cos\phi_1\sin^3\theta\cos\theta\sin\phi\cos\phi{}+3\sin\theta_1\cos^2\theta_1\cos\phi_1\sin^2\theta\cos^2\theta\cos\phi{}\br
& &
+\sin^3\theta_1\sin^3\phi_1\sin^4\theta\sin^3\phi{}+3\sin^2\theta_1\cos\theta_1\sin^2\phi_1\sin^3\theta\cos\theta\sin^2\phi{}\br
& &
+3\sin\theta_1\cos^2\theta_1\sin\phi_1\sin^2\theta\cos^2\theta\sin\phi{}+\cos^3\theta_1\sin\theta\cos^3\theta{}))\td{\theta}\td{\phi}\br
& = & 4\pi{}y_1z_1^2\pnc{,}
\end{eqnarray}
\begin{eqnarray}
I_6 & = & \int_{\theta{}=0}^{\pi}\int_{\phi=0}^{2\pi}x_1^3\sin^2\theta\cos\phi\td{\theta}\td{\phi}\br
& = &
\int_{\theta{}=0}^{\pi}\int_{\phi=0}^{2\pi}(y_1^3\sin^2\theta\cos\phi\br
& & +3y_1^2z_1(\sin\theta_1\cos\phi_1\sin^3\theta\cos^2\phi{}+\sin\theta_1\sin\phi_1\sin^3\theta\sin\phi\cos\phi{}+\cos\theta_1\sin^2\theta\cos\theta\cos\phi{})\br
& & +3y_1z_1^2(\sin^2\theta_1\cos^2\phi_1\sin^4\theta\cos^3\phi{}+2\sin^2\theta_1\sin\phi_1\cos\phi_1\sin^4\theta\sin\phi{}\cos^2\phi{}\br
& &
+2\sin\theta_1\cos\theta_1\cos\phi_1\sin^3\theta\cos\theta\cos^2\phi{}+\sin^2\theta_1\sin^2\phi_1\sin^4\theta\sin^2\phi\cos\phi\br
& &
+2\sin\theta_1\cos\theta_1\sin\phi_1\sin^3\theta\cos\theta\sin\phi\cos\phi{}+\cos^2\theta_1\sin^2\theta\cos^2\theta\cos\phi{})\br
& &
+z_1^3(\sin^3\theta_1\cos^3\phi_1\sin^5\theta\cos^4\phi{}+3\sin^3\theta_1\sin\phi_1\cos^2\phi_1\sin^5\theta\sin\phi\cos^3\phi{}\br
& &
+3\sin^2\theta_1\cos\theta_1\cos^2\phi_1\sin^4\theta\cos\theta\cos^3\phi{}+3\sin^3\theta_1\sin^2\phi_1\cos\phi_1\sin^5\theta\sin^2\phi\cos^2\phi{}\br
& &
+6\sin^2\theta_1\cos\theta_1\sin\phi_1\cos\phi_1\sin^4\theta\cos\theta\sin\phi\cos^2\phi{}+3\sin\theta_1\cos^2\theta_1\cos\phi_1\sin^3\theta\cos^2\theta\cos^2\phi{}\br
& &
+\sin^3\theta_1\sin^3\phi_1\sin^5\theta\sin^3\phi\cos\phi{}+3\sin^2\theta_1\cos\theta_1\sin^2\phi_1\sin^4\theta\cos\theta\sin^2\phi\cos\phi{}\br
& &
+3\sin\theta_1\cos^2\theta_1\sin\phi_1\sin^3\theta\cos^2\theta\sin\phi\cos\phi{}+\cos^3\theta_1\sin^2\theta\cos^3\theta\cos\phi{}))\td{\theta}\td{\phi}\br
& = & \left(4\pi{}y_1^2z_1+\frac{4\pi{}z_1^3}{5}\right)\sin\theta_1\cos\phi_1\pnc{,}
\end{eqnarray}
\begin{eqnarray}
I_7 & = &
\int_{\theta{}=0}^{\pi}\int_{\phi=0}^{2\pi}x_1^3\sin^2\theta\sin\phi\td{\theta}\td{\phi}\br
& = &
\int_{\theta{}=0}^{\pi}\int_{\phi=0}^{2\pi}(y_1^3\sin^2\theta\sin\phi\br
& & +3y_1^2z_1(\sin\theta_1\cos\phi_1\sin^3\theta\sin\phi\cos\phi{}+\sin\theta_1\sin\phi_1\sin^3\theta\sin^2\phi{}+\cos\theta_1\sin^2\theta\cos\theta\sin\phi{})\br
& & +3y_1z_1^2(\sin^2\theta_1\cos^2\phi_1\sin^4\theta\sin\phi\cos^2\phi{}+2\sin^2\theta_1\sin\phi_1\cos\phi_1\sin^4\theta\sin^2\phi{}\cos\phi{}\br
& &
+2\sin\theta_1\cos\theta_1\cos\phi_1\sin^3\theta\cos\theta\sin\phi\cos\phi{}+\sin^2\theta_1\sin^2\phi_1\sin^4\theta\sin^3\phi\br
& &
+2\sin\theta_1\cos\theta_1\sin\phi_1\sin^3\theta\cos\theta\sin^2\phi{}+\cos^2\theta_1\sin^2\theta\cos^2\theta\sin\phi{})\br
& &
+z_1^3(\sin^3\theta_1\cos^3\phi_1\sin^5\theta\sin\phi\cos^3\phi{}+3\sin^3\theta_1\sin\phi_1\cos^2\phi_1\sin^5\theta\sin^2\phi\cos^2\phi{}\br
& &
+3\sin^2\theta_1\cos\theta_1\cos^2\phi_1\sin^4\theta\cos\theta\sin\phi\cos^2\phi{}+3\sin^3\theta_1\sin^2\phi_1\cos\phi_1\sin^5\theta\sin^3\phi\cos\phi{}\br
& &
+6\sin^2\theta_1\cos\theta_1\sin\phi_1\cos\phi_1\sin^4\theta\cos\theta\sin^2\phi\cos\phi{}+3\sin\theta_1\cos^2\theta_1\cos\phi_1\sin^3\theta\cos^2\theta\sin\phi\cos\phi{}\br
& &
+\sin^3\theta_1\sin^3\phi_1\sin^5\theta\sin^4\phi{}+3\sin^2\theta_1\cos\theta_1\sin^2\phi_1\sin^4\theta\cos\theta\sin^3\phi{}\br
& &
+3\sin\theta_1\cos^2\theta_1\sin\phi_1\sin^3\theta\cos^2\theta\sin^2\phi{}+\cos^3\theta_1\sin^2\theta\cos^3\theta\sin\phi{}))\td{\theta}\td{\phi}\br
& = & \left(4\pi{}y_1^2z_1+\frac{4\pi{}z_1^3}{5}\right)\sin\theta_1\sin\phi_1\pnc{,}
\end{eqnarray}
and
\begin{eqnarray}
I_8 & = &
\int_{\theta{}=0}^{\pi}\int_{\phi=0}^{2\pi}x_1^3\sin\theta\cos\theta\td{\theta}\td{\phi}\br
& = &
\int_{\theta{}=0}^{\pi}\int_{\phi=0}^{2\pi}(y_1^3\sin\theta\cos\theta\br
& & +3y_1^2z_1(\sin\theta_1\cos\phi_1\sin^2\theta\cos\theta\cos\phi{}+\sin\theta_1\sin\phi_1\sin^2\theta\cos\theta\sin\phi{}+\cos\theta_1\sin\theta\cos^2\theta{})\br
& & +3y_1z_1^2(\sin^2\theta_1\cos^2\phi_1\sin^3\theta\cos\theta\cos^2\phi{}+2\sin^2\theta_1\sin\phi_1\cos\phi_1\sin^3\theta\cos\theta\sin\phi{}\cos\phi{}\br
& &
+2\sin\theta_1\cos\theta_1\cos\phi_1\sin^2\theta\cos^2\theta\cos\phi{}+\sin^2\theta_1\sin^2\phi_1\sin^3\theta\cos\theta\sin^2\phi\br
& &
+2\sin\theta_1\cos\theta_1\sin\phi_1\sin^2\theta\cos^2\theta\sin\phi{}+\cos^2\theta_1\sin\theta\cos^3\theta{})\br
& &
+z_1^3(\sin^3\theta_1\cos^3\phi_1\sin^4\theta\cos\theta\cos^3\phi{}+3\sin^3\theta_1\sin\phi_1\cos^2\phi_1\sin^4\theta\cos\theta\sin\phi\cos^2\phi{}\br
& &
+3\sin^2\theta_1\cos\theta_1\cos^2\phi_1\sin^3\theta\cos^2\theta\cos^2\phi{}+3\sin^3\theta_1\sin^2\phi_1\cos\phi_1\sin^4\theta\cos\theta\sin^2\phi\cos\phi{}\br
& &
+6\sin^2\theta_1\cos\theta_1\sin\phi_1\cos\phi_1\sin^3\theta\cos^2\theta\sin\phi\cos\phi{}+3\sin\theta_1\cos^2\theta_1\cos\phi_1\sin^2\theta\cos^3\theta\cos\phi{}\br
& &
+\sin^3\theta_1\sin^3\phi_1\sin^4\theta\cos\theta\sin^3\phi{}+3\sin^2\theta_1\cos\theta_1\sin^2\phi_1\sin^3\theta\cos^2\theta\sin^2\phi{}\br
& &
+3\sin\theta_1\cos^2\theta_1\sin\phi_1\sin^2\theta\cos^3\theta\sin\phi{}+\cos^3\theta_1\sin\theta\cos^4\theta{}))\td{\theta}\td{\phi}\br
& = & \left(4\pi{}y_1^2z_1+\frac{4\pi{}z_1^3}{5}\right)\cos\theta_1\pnc{.}
\end{eqnarray}

This gives a polarization
\begin{eqnarray}
\vc{P} & = &
\frac{(I_2-I_6,I_3-I_7,I_4-I_8)+O(\{y_1,z_1\}^4)}{I_1-I_5+O(\{y_1,z_1\}^4)}\br
& = &
\frac{(10y_1z_1-15y_1^2z_1-3z_1^3)(\sin\theta_1\cos\phi_1,\sin\theta_1\sin\phi_1,\cos\theta_1)+O(\{y_1,z_1\}^4)}{15y_1^2+5z_1^2-15y_1z_1^2+O(\{y_1,z_1\}^4)}\br
& = &
\frac{(10y_1z_1-15y_1^2z_1-3z_1^3)\vc{\hat{B_1}}+O(\{y_1,z_1\}^4)}{15y_1^2+5z_1^2}\left(1-\frac{5y_1z_1^2}{3y_1^2+z_1^2}+O(\{y_1,z_1\}^2)\right)^{-1}\br
& = &
\frac{2y_1z_1\vc{\hat{B_1}}}{3y_1^2+z_1^2}+\frac{(36y_1^2z_1^3-45y_1^4z_1-3z_1^5)\vc{\hat{B_1}}}{45y_1^4+30y_1^2z_1^2+5z_1^4}+O(\{y_1,z_1\}^2)\pnc{,}
\end{eqnarray}
where $\vc{\hat{B_1}}$ is a unit vector, in the direction of the
magnetic flux density in region 1.

\appendix

\section{GNU Free Documentation License}
\label{copying.data}

Version 1.1, March 2000\\

 Copyright \copyright\ 2000  Free Software Foundation, Inc.\\
     59 Temple Place, Suite 330, Boston, MA  02111-1307  USA\\
 Everyone is permitted to copy and distribute verbatim copies
 of this license document, but changing it is not allowed.

\subsection*{Preamble}

The purpose of this License is to make a manual, textbook, or other
written document ``free'' in the sense of freedom: to assure everyone
the effective freedom to copy and redistribute it, with or without
modifying it, either commercially or noncommercially.  Secondarily,
this License preserves for the author and publisher a way to get
credit for their work, while not being considered responsible for
modifications made by others.

This License is a kind of ``copyleft'', which means that derivative
works of the document must themselves be free in the same sense.  It
complements the GNU General Public License, which is a copyleft
license designed for free software.

We have designed this License in order to use it for manuals for free
software, because free software needs free documentation: a free
program should come with manuals providing the same freedoms that the
software does.  But this License is not limited to software manuals;
it can be used for any textual work, regardless of subject matter or
whether it is published as a printed book.  We recommend this License
principally for works whose purpose is instruction or reference.

\subsection{Applicability and Definitions}

This License applies to any manual or other work that contains a
notice placed by the copyright holder saying it can be distributed
under the terms of this License.  The ``Document'', below, refers to any
such manual or work.  Any member of the public is a licensee, and is
addressed as ``you''.

A ``Modified Version'' of the Document means any work containing the
Document or a portion of it, either copied verbatim, or with
modifications and/or translated into another language.

A ``Secondary Section'' is a named appendix or a front-matter section of
the Document that deals exclusively with the relationship of the
publishers or authors of the Document to the Document's overall subject
(or to related matters) and contains nothing that could fall directly
within that overall subject.  (For example, if the Document is in part a
textbook of mathematics, a Secondary Section may not explain any
mathematics.)  The relationship could be a matter of historical
connection with the subject or with related matters, or of legal,
commercial, philosophical, ethical or political position regarding
them.

The ``Invariant Sections'' are certain Secondary Sections whose titles
are designated, as being those of Invariant Sections, in the notice
that says that the Document is released under this License.

The ``Cover Texts'' are certain short passages of text that are listed,
as Front-Cover Texts or Back-Cover Texts, in the notice that says that
the Document is released under this License.

A ``Transparent'' copy of the Document means a machine-readable copy,
represented in a format whose specification is available to the
general public, whose contents can be viewed and edited directly and
straightforwardly with generic text editors or (for images composed of
pixels) generic paint programs or (for drawings) some widely available
drawing editor, and that is suitable for input to text formatters or
for automatic translation to a variety of formats suitable for input
to text formatters.  A copy made in an otherwise Transparent file
format whose markup has been designed to thwart or discourage
subsequent modification by readers is not Transparent.  A copy that is
not ``Transparent'' is called ``Opaque''.

Examples of suitable formats for Transparent copies include plain
ASCII without markup, Texinfo input format, \LaTeX~input format, SGML
or XML using a publicly available DTD, and standard-conforming simple
HTML designed for human modification.  Opaque formats include
PostScript, PDF, proprietary formats that can be read and edited only
by proprietary word processors, SGML or XML for which the DTD and/or
processing tools are not generally available, and the
machine-generated HTML produced by some word processors for output
purposes only.

The ``Title Page'' means, for a printed book, the title page itself,
plus such following pages as are needed to hold, legibly, the material
this License requires to appear in the title page.  For works in
formats which do not have any title page as such, ``Title Page'' means
the text near the most prominent appearance of the work's title,
preceding the beginning of the body of the text.

\subsection{Verbatim Copying}

You may copy and distribute the Document in any medium, either
commercially or noncommercially, provided that this License, the
copyright notices, and the license notice saying this License applies
to the Document are reproduced in all copies, and that you add no other
conditions whatsoever to those of this License.  You may not use
technical measures to obstruct or control the reading or further
copying of the copies you make or distribute.  However, you may accept
compensation in exchange for copies.  If you distribute a large enough
number of copies you must also follow the conditions in section 3.

You may also lend copies, under the same conditions stated above, and
you may publicly display copies.

\subsection{Copying in Quantity}

If you publish printed copies of the Document numbering more than 100,
and the Document's license notice requires Cover Texts, you must enclose
the copies in covers that carry, clearly and legibly, all these Cover
Texts: Front-Cover Texts on the front cover, and Back-Cover Texts on
the back cover.  Both covers must also clearly and legibly identify
you as the publisher of these copies.  The front cover must present
the full title with all words of the title equally prominent and
visible.  You may add other material on the covers in addition.
Copying with changes limited to the covers, as long as they preserve
the title of the Document and satisfy these conditions, can be treated
as verbatim copying in other respects.

If the required texts for either cover are too voluminous to fit
legibly, you should put the first ones listed (as many as fit
reasonably) on the actual cover, and continue the rest onto adjacent
pages.

If you publish or distribute Opaque copies of the Document numbering
more than 100, you must either include a machine-readable Transparent
copy along with each Opaque copy, or state in or with each Opaque copy
a publicly-accessible computer-network location containing a complete
Transparent copy of the Document, free of added material, which the
general network-using public has access to download anonymously at no
charge using public-standard network protocols.  If you use the latter
option, you must take reasonably prudent steps, when you begin
distribution of Opaque copies in quantity, to ensure that this
Transparent copy will remain thus accessible at the stated location
until at least one year after the last time you distribute an Opaque
copy (directly or through your agents or retailers) of that edition to
the public.

It is requested, but not required, that you contact the authors of the
Document well before redistributing any large number of copies, to give
them a chance to provide you with an updated version of the Document.

\subsection{Modifications}

You may copy and distribute a Modified Version of the Document under
the conditions of sections 2 and 3 above, provided that you release
the Modified Version under precisely this License, with the Modified
Version filling the role of the Document, thus licensing distribution
and modification of the Modified Version to whoever possesses a copy
of it.  In addition, you must do these things in the Modified Version:

\begin{itemize}

\item Use in the Title Page (and on the covers, if any) a title distinct
   from that of the Document, and from those of previous versions
   (which should, if there were any, be listed in the History section
   of the Document).  You may use the same title as a previous version
   if the original publisher of that version gives permission.
\item List on the Title Page, as authors, one or more persons or entities
   responsible for authorship of the modifications in the Modified
   Version, together with at least five of the principal authors of the
   Document (all of its principal authors, if it has less than five).
\item State on the Title page the name of the publisher of the
   Modified Version, as the publisher.
\item Preserve all the copyright notices of the Document.
\item Add an appropriate copyright notice for your modifications
   adjacent to the other copyright notices.
\item Include, immediately after the copyright notices, a license notice
   giving the public permission to use the Modified Version under the
   terms of this License, in the form shown in the Addendum below.
\item Preserve in that license notice the full lists of Invariant Sections
   and required Cover Texts given in the Document's license notice.
\item Include an unaltered copy of this License.
\item Preserve the section entitled ``History'', and its title, and add to
   it an item stating at least the title, year, new authors, and
   publisher of the Modified Version as given on the Title Page.  If
   there is no section entitled ``History'' in the Document, create one
   stating the title, year, authors, and publisher of the Document as
   given on its Title Page, then add an item describing the Modified
   Version as stated in the previous sentence.
\item Preserve the network location, if any, given in the Document for
   public access to a Transparent copy of the Document, and likewise
   the network locations given in the Document for previous versions
   it was based on.  These may be placed in the ``History'' section.
   You may omit a network location for a work that was published at
   least four years before the Document itself, or if the original
   publisher of the version it refers to gives permission.
\item In any section entitled ``Acknowledgements'' or ``Dedications'',
   preserve the section's title, and preserve in the section all the
   substance and tone of each of the contributor acknowledgements
   and/or dedications given therein.
\item Preserve all the Invariant Sections of the Document,
   unaltered in their text and in their titles.  Section numbers
   or the equivalent are not considered part of the section titles.
\item Delete any section entitled ``Endorsements''.  Such a section
   may not be included in the Modified Version.
\item Do not retitle any existing section as ``Endorsements''
   or to conflict in title with any Invariant Section.

\end{itemize}

If the Modified Version includes new front-matter sections or
appendices that qualify as Secondary Sections and contain no material
copied from the Document, you may at your option designate some or all
of these sections as invariant.  To do this, add their titles to the
list of Invariant Sections in the Modified Version's license notice.
These titles must be distinct from any other section titles.

You may add a section entitled ``Endorsements'', provided it contains
nothing but endorsements of your Modified Version by various
parties -- for example, statements of peer review or that the text has
been approved by an organization as the authoritative definition of a
standard.

You may add a passage of up to five words as a Front-Cover Text, and a
passage of up to 25 words as a Back-Cover Text, to the end of the list
of Cover Texts in the Modified Version.  Only one passage of
Front-Cover Text and one of Back-Cover Text may be added by (or
through arrangements made by) any one entity.  If the Document already
includes a cover text for the same cover, previously added by you or
by arrangement made by the same entity you are acting on behalf of,
you may not add another; but you may replace the old one, on explicit
permission from the previous publisher that added the old one.

The author(s) and publisher(s) of the Document do not by this License
give permission to use their names for publicity for or to assert or
imply endorsement of any Modified Version.

\subsection{Combining Documents}

You may combine the Document with other documents released under this
License, under the terms defined in section 4 above for modified
versions, provided that you include in the combination all of the
Invariant Sections of all of the original documents, unmodified, and
list them all as Invariant Sections of your combined work in its
license notice.

The combined work need only contain one copy of this License, and
multiple identical Invariant Sections may be replaced with a single
copy.  If there are multiple Invariant Sections with the same name but
different contents, make the title of each such section unique by
adding at the end of it, in parentheses, the name of the original
author or publisher of that section if known, or else a unique number.
Make the same adjustment to the section titles in the list of
Invariant Sections in the license notice of the combined work.

In the combination, you must combine any sections entitled ``History''
in the various original documents, forming one section entitled
``History''; likewise combine any sections entitled ``Acknowledgements'',
and any sections entitled ``Dedications''.  You must delete all sections
entitled ``Endorsements.''

\subsection{Collections of Documents}

You may make a collection consisting of the Document and other documents
released under this License, and replace the individual copies of this
License in the various documents with a single copy that is included in
the collection, provided that you follow the rules of this License for
verbatim copying of each of the documents in all other respects.

You may extract a single document from such a collection, and distribute
it individually under this License, provided you insert a copy of this
License into the extracted document, and follow this License in all
other respects regarding verbatim copying of that document.

\subsection{Aggregation With Independent Works}

A compilation of the Document or its derivatives with other separate
and independent documents or works, in or on a volume of a storage or
distribution medium, does not as a whole count as a Modified Version
of the Document, provided no compilation copyright is claimed for the
compilation.  Such a compilation is called an ``aggregate'', and this
License does not apply to the other self-contained works thus compiled
with the Document, on account of their being thus compiled, if they
are not themselves derivative works of the Document.

If the Cover Text requirement of section 3 is applicable to these
copies of the Document, then if the Document is less than one quarter
of the entire aggregate, the Document's Cover Texts may be placed on
covers that surround only the Document within the aggregate.
Otherwise they must appear on covers around the whole aggregate.

\subsection{Translation}

Translation is considered a kind of modification, so you may
distribute translations of the Document under the terms of section 4.
Replacing Invariant Sections with translations requires special
permission from their copyright holders, but you may include
translations of some or all Invariant Sections in addition to the
original versions of these Invariant Sections.  You may include a
translation of this License provided that you also include the
original English version of this License.  In case of a disagreement
between the translation and the original English version of this
License, the original English version will prevail.

\subsection{Termination}

You may not copy, modify, sublicense, or distribute the Document except
as expressly provided for under this License.  Any other attempt to
copy, modify, sublicense or distribute the Document is void, and will
automatically terminate your rights under this License.  However,
parties who have received copies, or rights, from you under this
License will not have their licenses terminated so long as such
parties remain in full compliance.

\subsection{Future Revisions of This License}

The Free Software Foundation may publish new, revised versions
of the GNU Free Documentation License from time to time.  Such new
versions will be similar in spirit to the present version, but may
differ in detail to address new problems or concerns. See
http://www.gnu.org/copyleft/.

Each version of the License is given a distinguishing version number.
If the Document specifies that a particular numbered version of this
License "or any later version" applies to it, you have the option of
following the terms and conditions either of that specified version or
of any later version that has been published (not as a draft) by the
Free Software Foundation.  If the Document does not specify a version
number of this License, you may choose any version ever published (not
as a draft) by the Free Software Foundation.

\subsection*{ADDENDUM: How to use this License for your documents}

To use this License in a document you have written, include a copy of
the License in the document and put the following copyright and
license notices just after the title page:

\begin{quote}

      Copyright \copyright\ YEAR  YOUR NAME.
      Permission is granted to copy, distribute and/or modify this document
      under the terms of the GNU Free Documentation License, Version 1.1
      or any later version published by the Free Software Foundation;
      with the Invariant Sections being LIST THEIR TITLES, with the
      Front-Cover Texts being LIST, and with the Back-Cover Texts being LIST.
      A copy of the license is included in the section entitled ``GNU
      Free Documentation License''.

\end{quote}

If you have no Invariant Sections, write ``with no Invariant Sections''
instead of saying which ones are invariant.  If you have no
Front-Cover Texts, write ``no Front-Cover Texts'' instead of
``Front-Cover Texts being LIST''; likewise for Back-Cover Texts.

If your document contains nontrivial examples of program code, we
recommend releasing these examples in parallel under your choice of
free software license, such as the GNU General Public License,
to permit their use in free software.

\bibliographystyle{abbrv}
\bibliography{journal-abbreviations,everything}

\end{document}